\documentclass[%
 reprint,
superscriptaddress,
 amsmath,amssymb,
 aps,
pra,
]{revtex4-1}

\usepackage{braket}
\usepackage{graphicx}
\usepackage{dcolumn}
\usepackage{bm}
\usepackage{hyperref}
\usepackage{multirow}
\usepackage{soul}
\hypersetup{colorlinks=true, citecolor=blue, urlcolor=blue, linkcolor=blue}

\begin{document}

\preprint{APS/123-QED}

\title{High-Rate $16$-node quantum access network based on passive optical network}

\author{Yan Pan}
 \email{These two authors contribute equally to this work.}
 \affiliation{%
 National Key Laboratory of Security Communication, Institute of Southwestern Communication, Chengdu 610041, China
}%

\author{Yiming Bian}
 \email{These two authors contribute equally to this work.}
 \affiliation{%
 State Key Laboratory of Information Photonics and Optical Communications, School of Electronic Engineering, Beijing University of Posts and Telecommunications, Beijing, 100876, China
}%

\author{Yang Li}
 \affiliation{%
 National Key Laboratory of Security Communication, Institute of Southwestern Communication, Chengdu 610041, China
}%

\author{Xuesong Xu}
 \affiliation{%
 State Key Laboratory of Information Photonics and Optical Communications, School of Electronic Engineering, Beijing University of Posts and Telecommunications, Beijing, 100876, China
}%

\author{Li Ma}
 \affiliation{%
 National Key Laboratory of Security Communication, Institute of Southwestern Communication, Chengdu 610041, China
}%

\author{Heng Wang}
 \affiliation{%
 National Key Laboratory of Security Communication, Institute of Southwestern Communication, Chengdu 610041, China
}%

\author{Yujie Luo}
 \affiliation{%
 National Key Laboratory of Security Communication, Institute of Southwestern Communication, Chengdu 610041, China
}%

\author{Jiayi Dou}
 \affiliation{%
 State Key Laboratory of Information Photonics and Optical Communications, School of Electronic Engineering, Beijing University of Posts and Telecommunications, Beijing, 100876, China
}%

\author{Yaodi Pi}
 \affiliation{%
 National Key Laboratory of Security Communication, Institute of Southwestern Communication, Chengdu 610041, China
}%

\author{Jie Yang}
 \affiliation{%
 National Key Laboratory of Security Communication, Institute of Southwestern Communication, Chengdu 610041, China
}%
 \affiliation{%
 State Key Laboratory of Information Photonics and Optical Communications, School of Electronic Engineering, Beijing University of Posts and Telecommunications, Beijing, 100876, China
}%

\author{Wei Huang}
 \affiliation{%
 National Key Laboratory of Security Communication, Institute of Southwestern Communication, Chengdu 610041, China
}%

\author{Song Yu}%
 \affiliation{%
 State Key Laboratory of Information Photonics and Optical Communications, School of Electronic Engineering, Beijing University of Posts and Telecommunications, Beijing, 100876, China
}%

\author{Stefano Pirandola}
 \affiliation{%
 Department of Computer Science, University of York, York YO10 5GH, United Kingdom
}%

\author{Yichen Zhang}%
 \affiliation{%
 State Key Laboratory of Information Photonics and Optical Communications, School of Electronic Engineering, Beijing University of Posts and Telecommunications, Beijing, 100876, China
}%

\author{Bingjie Xu}
 \affiliation{%
 National Key Laboratory of Security Communication, Institute of Southwestern Communication, Chengdu 610041, China
}%

\date{\today}
\begin{abstract}
    Quantum key distribution can provide  {information-theoretical} secure communication, which is now heading towards building the quantum secure network for real-world applications. In most built quantum secure networks, point-to-multipoint (PTMP) topology is one of the most popular schemes, especially for quantum access networks. However, due to the lack of custom protocols with high secret key rate and compatible with classical optical networks for PTMP scheme, there is still no efficient way for a high-performance quantum access network with a multitude of users. Here, we report an experimental demonstration of a high-rate 16-nodes quantum access network based on passive optical network, where a high-efficient coherent-state PTMP protocol is novelly designed to allow independent secret key generation between one transmitter and multiple receivers concurrently. Such accomplishment is attributed to  {a well-designed real-time shot-noise calibration method, a series of advanced digital signal processing algorithms and a flexible post-processing strategy with high success probability}. Finally, the experimental results show that the average secret key rate is around 2.086 Mbps between the transmitter and each user, which is two orders of magnitude higher than previous demonstrations. With the advantages of low cost {,} excellent compatibility, and wide bandwidth, our work paves the way for building practical PTMP quantum access networks, thus constituting an important step towards scalable quantum secure networks.
\end{abstract}

\maketitle

Quantum key distribution (QKD) \cite{bennet1984quantum, AdvInQC, PTPQKDRMV2020} allows secret key generation between two distant parties secured by the fundamental laws of physics. It can be realized by encoding the secret information on quantum states within a finite or infinite Hilbert space, corresponding to the discrete variable (DV) and continuous variable (CV) protocols \cite{GG02Nature,GaussianQuantumInformation,lam2013continuous} respectively. The DV-QKD has experienced a long period of development and can support a rather long distance transmission distance \cite{PTPQKDTFNat2018, 800kmTFQKD}, while CV-QKD~\cite{CVReV2023} is advantageous in the compatibility with classical optical communications and high secret key rate (SKR) within metropolitan distances \cite{CvExp80km2012,CVMDIYork,CVQKD50km,CvExpSOICVQKD2019,CvExp202kmPRL,CVNC2022}.
In pace with the maturity of the point-to-point links, QKD is developing towards networking.
It has been realized from the metropolitan-area network \cite{QCnetVienna2009, QCNetToyoko, CambridgeQN, HuweiQN} to the large scale wide-area network \cite{Micius,QCnetNature2021}, even with space-to-ground links \cite{PTPQKDSatellite2017,QCnetNature2021,Micius}.

\begin{figure*}
    \centering
    \includegraphics[width=18 cm]{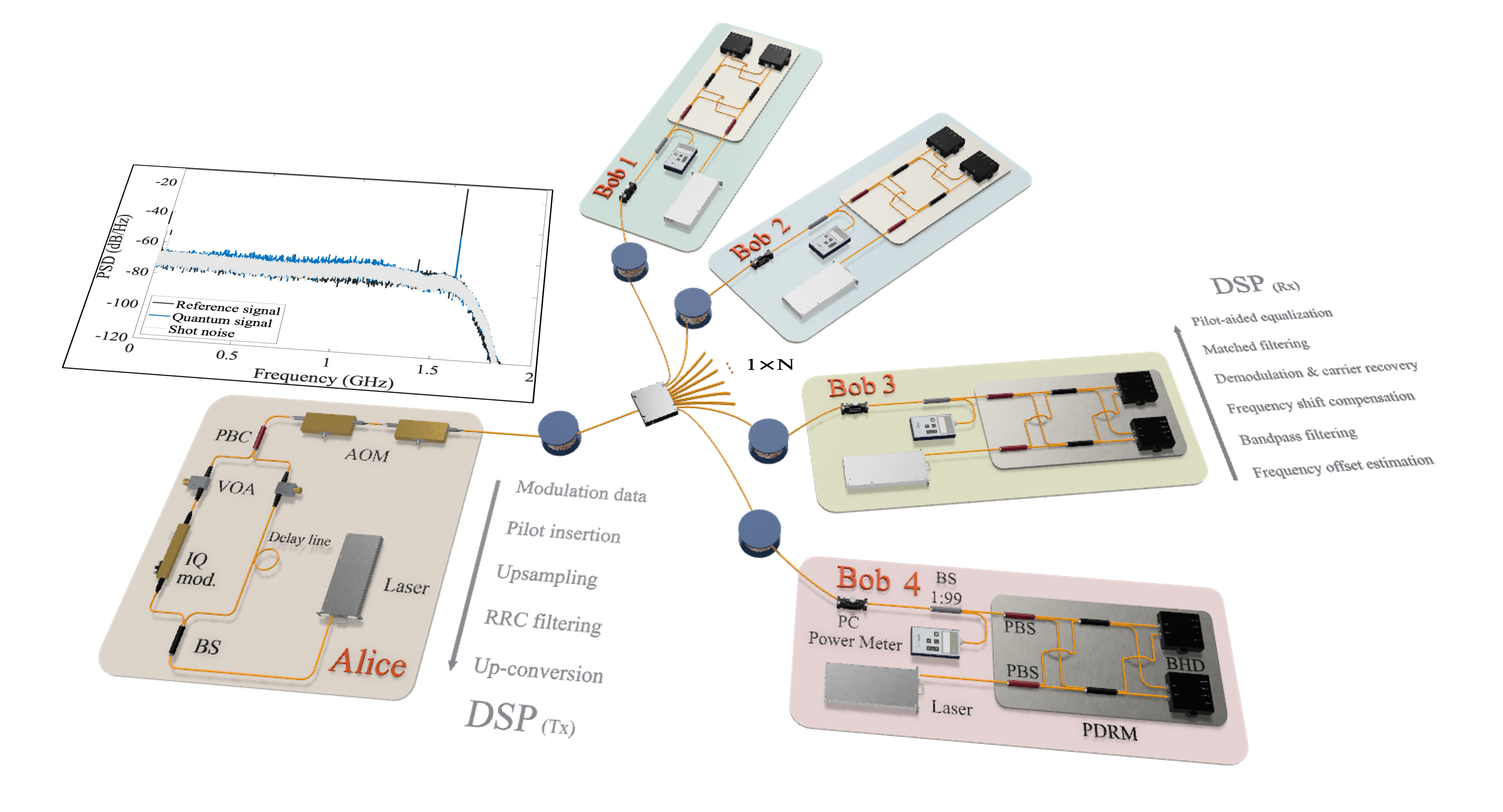}
    \caption{\textbf{Experimental setup of 16-node quantum access network based on passive optical network.} CW laser, continuous-wave laser; AWG, arbitrary waveform generator; IQ modulator, In-phase/quadrature modulator; ODL, optical delay line; VOA, variable optical attenuator; AFG, arbitrary function generator; PC, polarization controller; PBC, polarization beam combiner; AOM, acousto-optic modulator; SSMF, standard single mode fiber; PDRM, polarization diversity receiver module; DSO, digital storage oscilloscope. }\label{Fig: Structure}
\end{figure*}

As the ``last mile'' of the QKD network, quantum access network is of significant importance for the large-scale application of QKD. As early as 1997, a multi-user QKD scheme on a downstream passive optical fibre network was proposed ~\cite{townsend1997quantum}. This implementation requires a single photon detector for each user, which is difficult be large-scale applied. An upstream quantum access network is proposed in 2013, where multi-users transmit quantum signals to a common receiver~\cite{frohlich2013quantum}. In this case, the cost is controllable, but the SKR is completely limited by the receiver. Furthermore, based on the downstream or upstream architecture, various schemes of QKD network have been proposed and deployed  {\cite{frohlich2013quantum,46node,Huang2020experimental,Wang2021Practical,Xu2023Round}}. However, the maximum SKR of all existing works is less than Mbps. This is mainly due to that no efficient protocol can natively support the secure connections of multiple users. Based on the point-to-point QKD protocols, even the most advanced existing metropolitan and access networks need relays, multiplexing technologies or simply building multiple QKD links to access multiple users. This leads to a complex network with limited load capacity quantum access networks. Therefore, an efficient method to build a quantum access network for multitude of users to access to the QKD infrastructure is still missing.

Here, we experimentally demonstrated a high-performance downstream quantum access network using coherent states based on a passive optical network (PON), which can access up to $16$ end users.
Since high-bandwidth commercial components can be used for quantum state preparation and measurement, and existing PON infrastructure can be used as the quantum channel, the investigated scheme inherits the advantages of CV-QKD and passive optical network (i.e., low-cost, excellent compatibility and wide-bandwidth). Thus, a better connectivity and scalability can be provided for quantum access network.
Meanwhile, four innovative techniques are adopted to achieve a considerable system performance, including the advanced point-to-multipoint (PTMP) protocol that supports high SKR for multiple users simultaneously, one-time shot-noise unit calibration within real-time, signal-to-noise enhancement of pilot symbols with time-domain superposition and high-performance self-adaptive post-processing. In this case, the average SKR of the investigated quantum access network can reach to $2.086$~Mbps when the quantum signal transmission over $1 \times 16$ splitter and $6$~km standard single mode fiber, which results in $2$~orders of magnitude of enhancement compared with states of the art works. Moreover, experimental results in the case of 1$\times$4 and 1$\times$8 splitters with different transmission distances (15 and 30 km) are also given for comparison. These results show the proposed scheme is a promising way of building high-rate, large-scale and cost-effective QKD network.

~\\
\textbf{Results}
~\\
\textbf{Point-to-multipoint protocol for network}
~\\
A multiuser CV-QKD protocol is developed to support the PTMP quantum access network.
The prepare-and-measure scheme is briefly introduced as below: The Gaussian modulated coherent state prepared by Alice is sent to multiple Bobs through an untrusted general broadcast channel for independent heterodyne detections.
Then, post processing, including parameter estimation, error correction and privacy amplification, is performed.
The untrusted general broadcast channel means that the eveasdropper Eve can fully control the channel, including the loss, noise and structure. The only requirement is that, the state prepared by Alice can be received by Bobs after the influence of the channel.
Since the average photon number of coherent states is larger than one, each prepared quantum state can be received by all Bobs, which results in a multi-party system, and can be described by $\rho_{AB_1B_2...B_N}$.

The secret key rate {$K_N$} between Alice and Bob $N$ {is computed in reverse reconciliation and given by the asymptotic formula}
\begin{equation}
    \label{SKR}
    K_N = \beta I\left({A}:{B}_N\right)-\max{\left\{\max_{i \neq N}{I({B}_N:{B}_i)},\chi_{{B}_N{E}}\right\}},
\end{equation}
where $\beta$ is the reconciliation efficiency, $I\left({A}:{B}_N\right)$ is the mutual information between Alice and Bob $N$, $\chi_{{B}_N{E}}$ is Eve's Holevo information on Bob $N$, and $I({B}_N:{B}_i)$ is the mutual information between Bob $N$ and the generic Bob $i$. In this formulation, the key point is that we need to remove not only the information that Eve may have stolen by attacking the channel between Alice and Bob $N$ but also any residual information that the other Bobs may have about Bob $N$, so the key is secret and also independent for each Bob. We assume that the various Bobs are trusted, i.e., they do not cooperate with Eve and/or between themselves in the attempt to eavesdrop on Bob $N$. Under typical experimental conditions, the contribution of Eve's Holevo bound is usually larger than the residual Bob's correlations, meaning that the effective key rate may collapse to the usual point-to-point expression. Security analysis is detailed in Methods and Supplementary Materials.

\begin{figure}[t]
    \centering
    \includegraphics[width=9 cm]{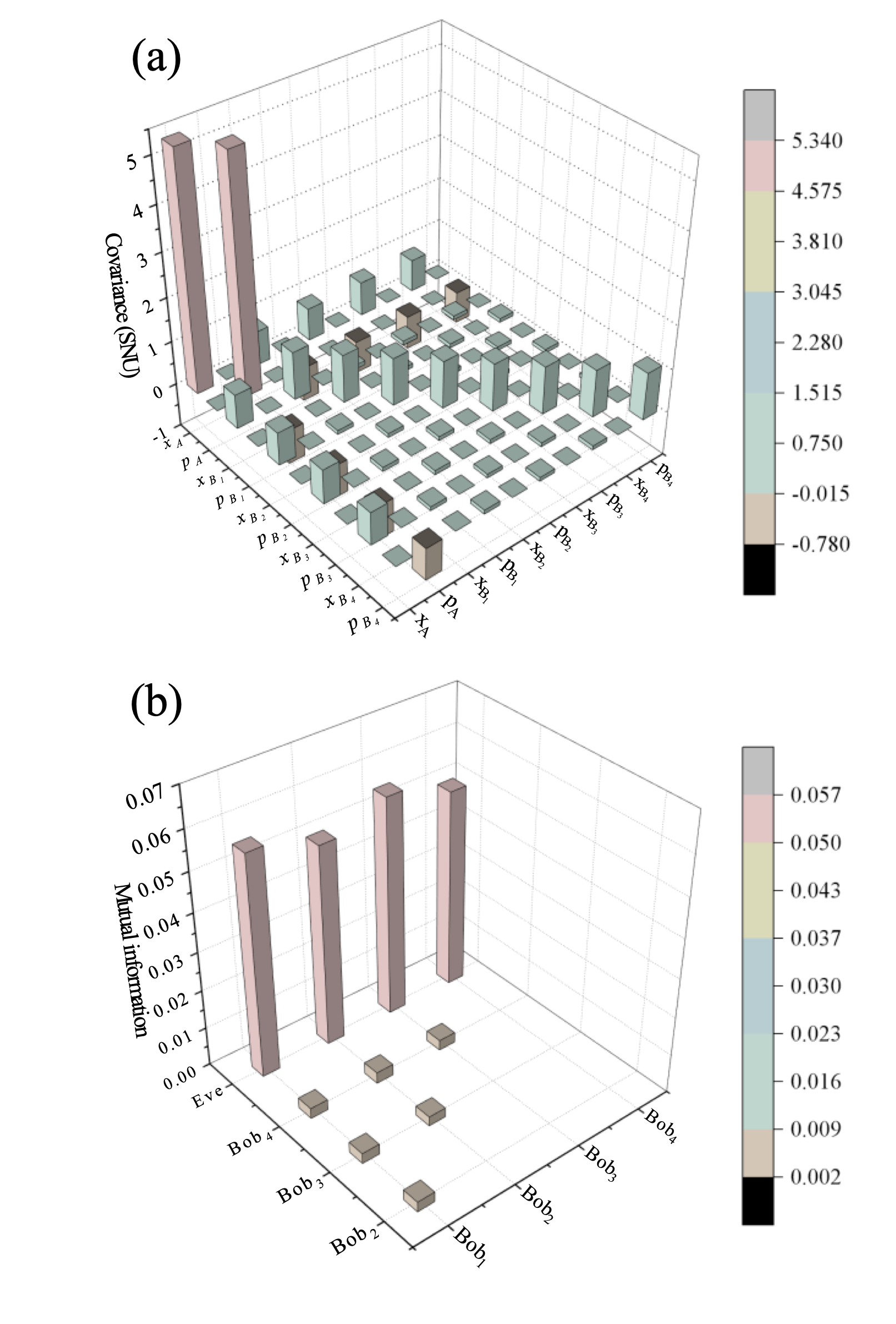}
    \caption{\textbf{Covariance matrix and mutual information.} (a) The covariance matrix $\gamma_{AB_1B_2B_3B_4}$. (b) The classical mutual information between different trusted receiver modes, and the Holevo bound between Eve and each receiver mode.}
    \label{Fig: Mat}
\end{figure}

~\\
\textbf{Experimental setup}
~\\
The experimental setup of the CV-QKD access network is shown in Fig.~\ref{Fig: Structure}. At Alice's site, a continuous-wave laser (NKT Photonic Basik) with a linewidth of $<100$~Hz is used as the optical carrier, and the wavelength is set to $1550.12$~nm. {The} light is launched into a beam splitter (BS) and split into two branches. One branch of the light is modulated with Gaussian signals by an $23$~GHz In-phase/quadrature (IQ) modulator (Fujitsu FTM7962EP). The $x$ and $p$ quadrature of Gaussian signals with a $750$~MHz frequency shift is generated by a two-channel arbitrary waveform generator that works at $30$~GSa/s (Keysight M8195A) for driving the IQ modulator. Here, the IQ modulator is working at optical single sideband modulation with carrier suppression, {realized} by an automatic bias controller.

{The} repetition frequency of two electrical Gaussian signals is set to $1$~GHz, and a digital root-raised cosine pulse shape filter with a roof-off factor of $0.3$ is performed to the baseband Gaussian signals. Meanwhile, a deterministic sequence of QPSK pilot symbols is interleaved in time with the quantum signals for training the optical channel impairments by the digital signal processing algorithms.
Notably, the amplitude of pilot symbols and quantum signals is equivalent, and no electrical amplifiers are used to avoid amplification noise and modulation nonlinearity.

{A} variable optical attenuator (VOA$_1$) is used to control the modulation variance $V_A$. Another branch of the light is used as the reference optical signal, and the reference path mainly consists of optical delay line and VOA$_2$. {Quantum} and pilot signals are combined by a polarization beam combiner. Therefore, the optical weak quantum and high-power reference signal are multiplexed by the dimension of polarization and frequency to suppress the crosstalk.
After that, two acousto-optic modulators (AOM) are used to control the on-off of multiplexed optical signals for real-time shot-noise unit (SNU) calibration. During SNU calibration, each AOM can bring $50$~dB extinction ratio, which suppresses the signal power to  {about} $-150$~dBm with two AOMs that has negligible effects on SNU calibration.

The transmission link consists of a $5/10/25$ km standard single mode fiber, an $1\times N$ optical power splitter, and $N$ segments of shorter fiber. At Bob's site, $4$ independent receivers are used for simulating different users. Here, the key parameters (such as transmissivity of transmission links and quantum efficiency of detectors) of the investigated 4 links are chosen to be different which is closer to practical application scenarios. In each receiver, an independent running continuous-wave laser with a linewidth of $<100$~Hz is used as the local oscillator, and the center wavelength is set to be about $1.55$~GHz shift from the laser at Alice's site. The state of polarization of the received optical signal is controlled by a polarization controller.

{To} avoid Eve's attack on SNU calibration, a sensitive optical power meter and a 1:99 beam splitter are used to monitor the optical power.
Then, the optical signal and local oscillator are coherently detected by a polarization diversity receiver module, which consists of a polarization beam splitter, a beam splitter, two polarization-maintaining optical couplers, and two balanced photo-detectors. Here, the $3$~dB bandwidth, responsibility, and gain of the detectors are $1.6$~GHz, $0.95$~A/W, and $3.0 \times 10^4$~V/A, respectively.

{Finally}, the received electrical signals are digitalized by a digital storage oscilloscope (Keysight MXR608A) working at 4 Gsa/s, and  offline DSP is performed for raw  {data} demodulation. Details of the DSP are shown in the following part of Methods. Notably, a time-domain superposition algorithm is designed to enhance the signal-to-noise ratio (SNR) of the deterministic training sequence, which determines the equalization accuracy of the quantum signal. Using the proposed algorithm, it is not necessary to drastically increase the amplitude of the training sequence at {Alice's} site, thereby avoiding the accuracy degradation of the quantum signal when generated/sampled by the limited resolution of DAC/ADC and transmission crosstalk caused by the high training sequence signal. Meanwhile, the SNR of training sequences can be flexibly controlled to meet the requirements of different links. More details about the time-domain superposition algorithm {are} in the Methods and Supplementary Material.

\begin{table*}[htbp]
    \centering
    \caption{Comparison of different quantum access network networks. $^*$ downstream network scheme. }
    \begin{ruledtabular}

    \begin{tabular}{ccccccc}
    \textbf{Literature}                      & \textbf{Connection}            & \textbf{Max. capacity} & \textbf{\begin{tabular}[c]{@{}c@{}}Distance \\ (km)\end{tabular}} & \textbf{\begin{tabular}[c]{@{}c@{}}Loss \\ (dB)\end{tabular}} & \textbf{\begin{tabular}[c]{@{}c@{}}Protocol\end{tabular}} & \textbf{\begin{tabular}[c]{@{}c@{}}SKR \\ (kbps)\end{tabular}}       \\    \hline
    \textit{Chen et al.}~\cite{46node}                     & Directly connected             & /                      & 18                                                            & /                                                             & DV                                                                  & 49.5                                                                 \\    \hline
    \multirow{2}{*}{\textit{Frohich et al.}~\cite{frohlich2013quantum}} & DWDM                           & 8                      & 16.2                                                          & 2.5                                                           & DV                                                                 & 303                                                                  \\
                                             & Beam splitter                  & 8                      & 16.2                                                          & 14                                                            & DV                                                                 & 47.5                                                                 \\    \hline
    \textit{Wang et al.}~\cite{Wang2021Practical}                     & Beam splitter $^*$                 & 16                     & 21                                                            & 22                                                            & DV                                                                 & 1.5                                                                  \\    \hline
    \textit{Huang et al.} ~\cite{Huang2020experimental}                   & Beam splitter                  & 2                      & 12.3                                                          & /                                                             & CV                                                                 & 22.19                                                                \\    \hline
    \textit{Xu et al.}~\cite{Xu2023Round}                       & Beam splitter                  & 8                      & 30                                                            & /                                                             & CV                                                                   & 0.82                                                                 \\    \hline
    \multirow{5}{*}{\textit{Our work}}       & \multirow{5}{*}{Beam splitter $^*$} & \multirow{2}{*}{4}     & 15                                                            & 10.9                                                          & CV                                                                & 12050                                                                 \\
                                             &                                &                        & 30                                                            & 12.8                                                          & CV                                                                & 4237                                                                 \\
                                             &                                & \multirow{2}{*}{8}     & 6                                                             & 11.3                                                          & CV                                                                & 7440                                                                 \\
                                             &                                &                        & 15                                                            & 14.2                                                          & CV                                                                & 3303                                                                 \\
                                             &                                & 16                     & 6                                                             & 15.1                                                          & CV                                                                & 2087
    \end{tabular}

    \end{ruledtabular}
    \label{Table: Results}
\end{table*}

Once each Bob has measured the states sent from Alice, every two parties postprocess their data to generate a secret key, via parameter estimation, information reconciliation, error correction, and privacy amplification.
In the parameter estimation procedure, the covariance matrix $\gamma_{AB_{1}B_{2}...B_{N}}$ is calculated as shown in Fig.~\ref{Fig: Mat}~(a), based on which the SKR between $A$ and each $B_i$ can be calculated as in Eq. \ref{SKR}.
High efficient error  {correction} is crucial for the performance of the investigated system.
Due to the inevitable difference in link loss and channel disturbance, the SNRs for each user are different and under slow fluctuation. Moreover, the modulation variance is hard to be optimized simultaneously for different QKD links, where one needs to continuously realize high-performance error-correction for each user with different and varying SNRs simultaneously, that is much more complex and challenging than the case in regular one-way point-to-point CV-QKD scenario. Fortunately, the SNR for each user under a specific fiber channel usually lies in a typical range, which can be calibrated priorly. Therefore, an efficient and practical error-correction method for PTMP CV-QKD protocol is {realized} based on the classification of SNRs and trusted noise addition technique.
In our experiment, 12 high-performance error-correction matrices are designed, which support stable and efficient error-correction for raw data with varying SNR of $0.041 - 0.048$. The optimal reconciliation efficiencies are 92.3\%, 92.6\%, 92.3\%, and 92.0\% for the investigated 4 users in the PTMP CV-QKD network, respectively. Details of post processing can be found in Supplementary Materials.


~\\
\textbf{Performance of the network}
~\\
The correlation between different Bobs can be defined by classical mutual information, denoted as $I_{B_{i}B_{j}}$. In the 16-node network, different Bobs' correlation is shown in Fig.~\ref{Fig: Mat}~(b). Compared with the upper bound of Eve's knowledge on each Bob ($\chi_{B_{i}E}$), $I_{B_{i}B_{j}}$ is significantly lower. Therefore, in PTMP scenario, {the correlations} between the legitimate end users is not the limiting factor of SKR. {This means that the performance of the PTMP scenario can be close to that of the point-to-point case.}

\begin{figure}[b]
    \centering
    \includegraphics[width=8 cm]{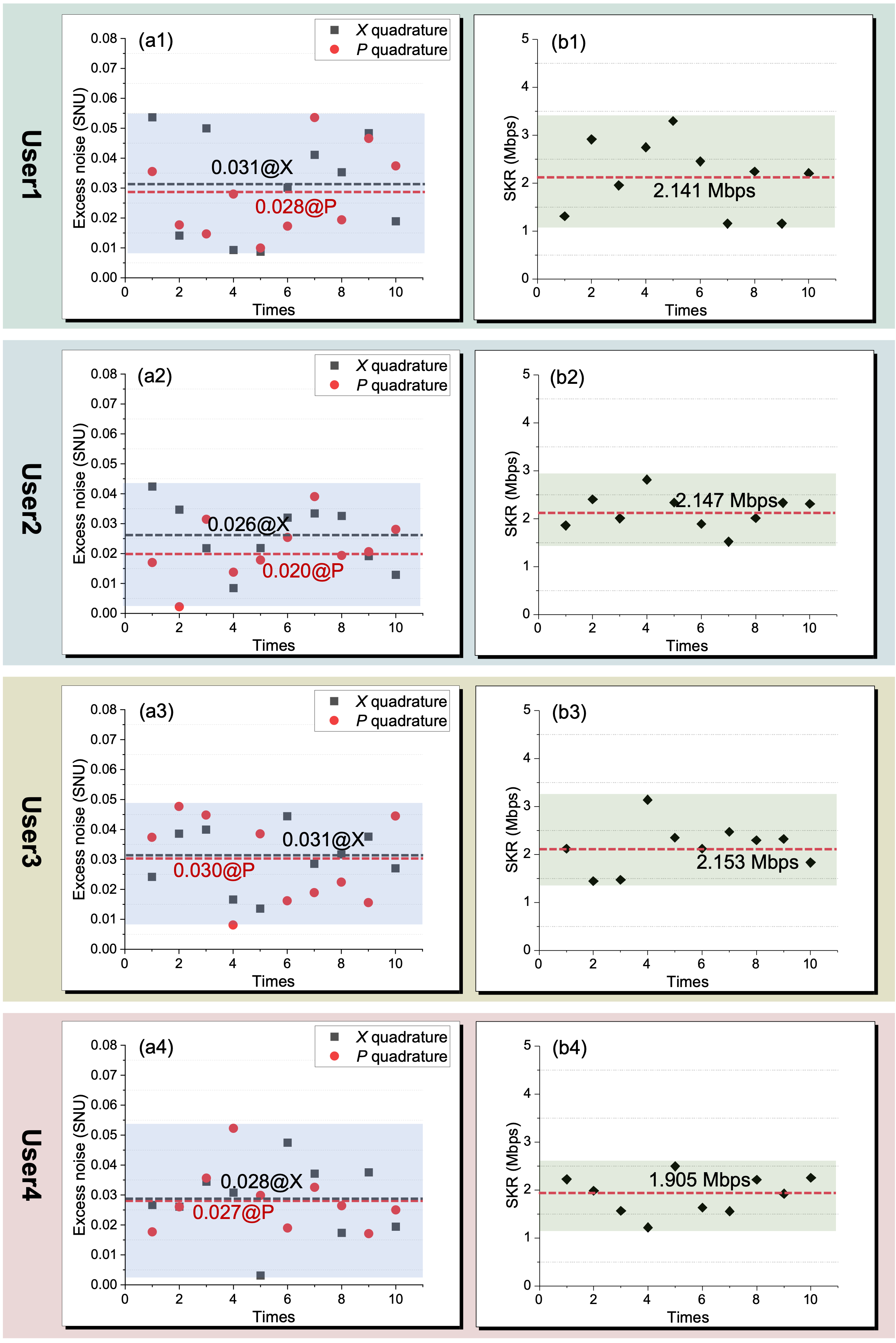}
    \caption{\textbf{The experimental results of $1 \times 16$ quantum access network for 4 users. }(a1-a4), Estimated excess noise with block size of $2.56 \times 10^8$. (b1-b4), Calculated secret key rate performance and numerical simulations. }\label{Fig: Results1}
\end{figure}

\begin{figure}
    \centering
    \includegraphics[width=8 cm]{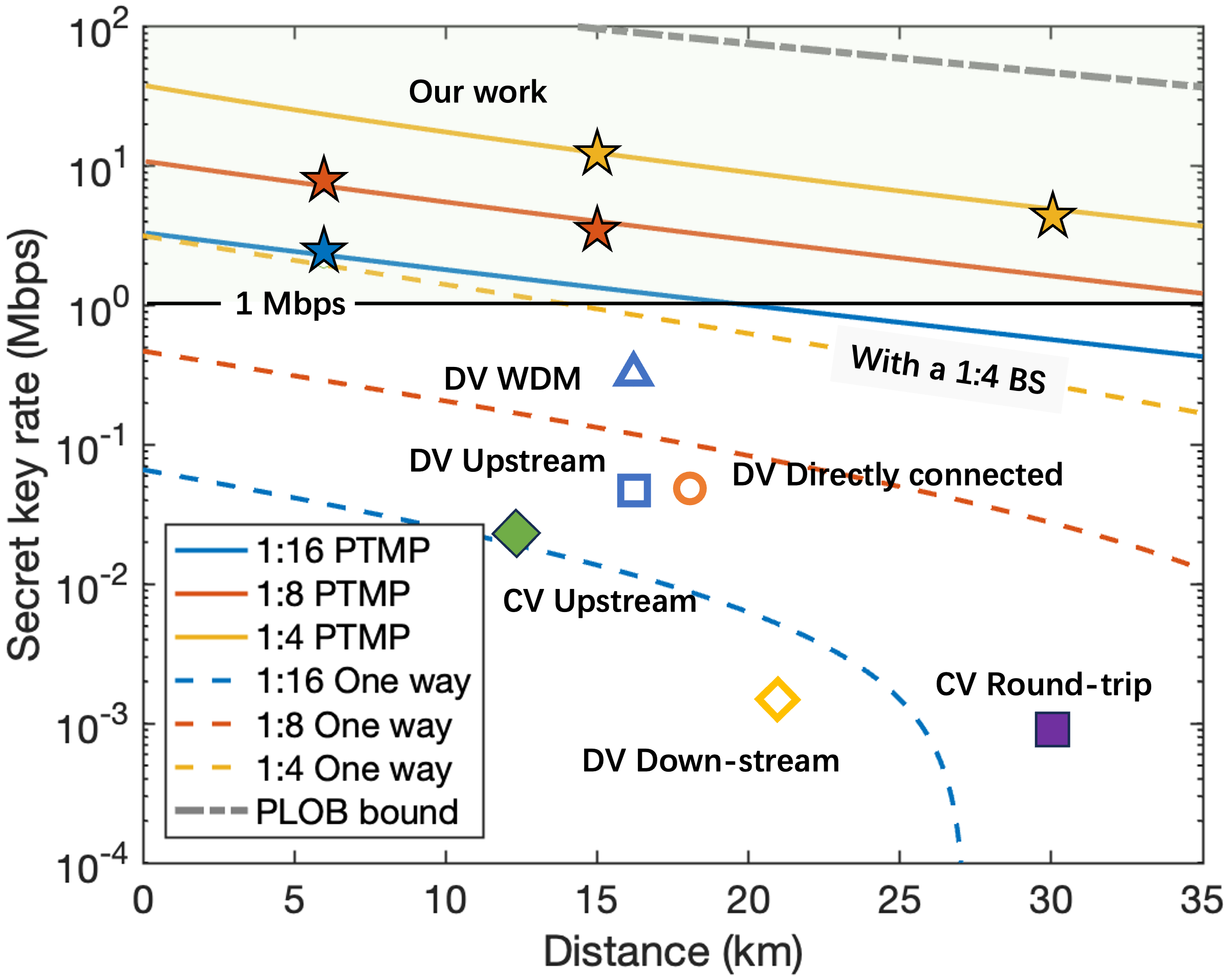}
    \caption{\textbf{Experimental key rates and numerical simulations. }The five five-pointed stars correspond to the experimental results at different fiber lengths with $1 \times 4$, $1 \times 8$, and $1 \times 16$ splitters. The blue, red, and yellow solid curve are the numerical simulation of the key rate which is computed starting from the experimental parameters with $1 \times 16$, $1 \times 8$, and $1 \times 4$ splitters, respectively. For comparison, we also show previous state-of-the-art quantum access network networks with discrete variable~\cite{46node,frohlich2013quantum,Wang2021Practical} and continuous variable~\cite{Huang2020experimental,Xu2023Round}. The PLOB bound is also given \cite{TheorPLOB2017}.}\label{Fig: Results2}
\end{figure}

Table~\ref{Table: Results} shows the key parameters and measured average results for tested 4 users of the 16-nodes quantum access network system. 
The detection efficiency including trusted insertion loss of the users are 0.71, 0.63, 0.63, and 0.75, and the link loss for the 4 users are 15.15, 14.95, 14.62, and 15.77 dB, respectively. In addition, the modulation variance is set to 4.3 SNU, the ratio of {the} training sequence is set to 20 \%, the reconciliation efficiency is about 92 \%. In this case, the estimated excess noise and SKR for the tested 4 users are shown in Fig.~\ref{Fig: Results1}. Here, every test dot is estimated by randomized 40 frames of data. The block size of every frame is $6.4 \times 10^6$, and the total block size for parameter estimation is $2.56 \times 10^8$.

{The} time of data acquisition is about 15 seconds for every frame. {In Fig.~\ref{Fig: Results1} ($a1 - a4$), we show the measured excess noises, with the dashed lines being the average values for each user. The average excess noise of the 4 users} is 0.031, 0.026, 0.031, and 0.028 SNU for {the} X quadrature, and 0.028, 0.020, 0.030, and 0.027 for {the} P quadrature, respectively. Meanwhile, as can be seen from Fig.~\ref{Fig: Results1} ($a1 - a4$), the X and P quadratures have a similar fluctuation of excess noise, in the range between 0 and 0.055 SNU. The fluctuation is mainly caused by detection noise and high link loss (16 dB).

The calculated SKR is shown in Table \ref{Table: Results} and Fig.~\ref{Fig: Results1} ($b1 - b4$), and the average SKRs are  {2.141, 2.147, 2.153, and 1.905 Mbps} for the tested 4 users, respectively. The fluctuation of the secret key rate is mainly due to the quantum signal recovery deviation and detection noise. Meanwhile, the SKR fluctuation of user 1 is wider, which is consistent with its stronger excess noise fluctuation. Fortunately, all the test results are $\geq$ 1 Mbps, and the average SKR is considerable.

The average SKR is compared with the state of art works in Fig.~\ref{Fig: Results2}. In Fig.~\ref{Fig: Results2}, the average asymptotic SKR of the investigated system is 12.050 Mbps, 4.237 Mbps, 7.440 Mbps, 3.303 Mbps and 2.086 Mbps, which is respectively achieved in the case of transmission over $1 \times 4$ splitter with 15 km and 30 km fiber link, $1 \times 8$ splitter with 6 km and 15 km fiber link, and $1 \times 16$ splitter with 6 km fiber link. Compared with the existing networks using point-to-point protocols (47.5 kbps @ 16.2 km~\cite{frohlich2013quantum} and 49.5 kbps @ 18 km~\cite{46node}), our results get an enhancement of 2 orders of magnitude. Even in comparison with results based on wavelength-dense-division multiplexing scheme (303 kbps @ 16.2 km~\cite{frohlich2013quantum}), which means less loss but much more complex and expensive system for the QKD network, our results still have an enhancement higher than an order of magnitude. Meanwhile, in the case of $1 \times 4$ splitter, the theoretical SKR curve by using point-to-point protocol is given for reference in Fig.~\ref{Fig: Results2}, and the SKR is more significantly improved for longer distances. Additional experiments results can be found in Supplementary Materials.

~\\
\textbf{Discussion}
~\\
In this work, a quantum access network with Mbps level SKR is experimentally demonstrated.  
For practical applications, the network channel in this experiment is completely compatible with the access network in classical optical communications, without the requirement of deploying additional fibers, which is the most concerned issue in optical access networks.
The downstream network scheme using a coherent state transmitter and multiple coherent receivers significantly promotes the scalable deployments. The receiver has the same structure as the coherent receiver in classical optical communications, which can be integrated on chip using photonic integrated circuit techniques and be suitable for cost-effective manufacturing in large scale \cite{CvExpSOICVQKD2019,wang2020integrated,CVReV2023}.

This scheme can be further implemented in the entanglement distribution network and the quantum internet, in which a continuous-variable source in a network node can be used to support multiple receivers owned by different end users, through a point-to-multipoint fiber channel with a beamsplitter scheme. It provides a possible way to construct and analysis the quantum state of the overall network, $\rho_{AB_1B_2...B_N}$, where $A$ is the mode of the source and $B_i$ is the mode of different users. In this way, the topology of a quantum internet can be simplified, where massive point-to-point links can be replaced by a point-to-multipoint network compatible with classical network facilities. It can be expected that the demonstrated scheme could be a favorable candidate for the end-user access in the next-generation quantum internet.

~\\
\textbf{Methods}
~\\
\textbf{Experiments details}
~\\
Key procedures of the experiment:
1) Optical Gaussian signal generation. In the experiment, IQ modulator is used for optical Gaussian signal generation by modulating X/P quadarature of Gaussian signal on the optical field directly. Here, the random bits are generated by a quantum random number generator, and converted in two independent sequences of decimal numbers every 16 bits. In this case, two sets of numbers satisfying independent uniform distributions are obtained, and named as $U_1$ and $U_2$. Then, according to the Box-Muller transform equation, the X/P quadrature of Gaussian signal can be calculated as $X=U_1\times cos(2\pi U_2)$ and $P=U_1\times sin(2\pi U_2)$. Meanwhile, the three-sigma truncation of X and P should be performed, and then the required signals X' and P' are obtained. After upsampling, root-raised cosine filering, and up-conversion for X' and P', the processed Gaussian signals are loaded onto the IQ modulator by AWG, and the optical Gaussian signal can be obtained.
2) Modulation variance (i.e. $V_A$) setting. According to simulations of $V_A$ optimization, $V_A$ is set to 4.32 SNU in the experiments. As a result, the transmitting power of the quantum signals can be obtained, which should be set to -65.5 dBm by adjusting VOA1.
3) Quantum efficiency calibration of PDRMs. The quantum efficiency of PDRMs is mainly affected by the insertion loss of optical passive components (i.e. PBS, optical couplers) and the quantum efficiency of BPDs in the PDRMs. Here, assuming the reliable responsivity (i.e. R) of BPDs, their quantum efficiency can be obtained as $\eta=1240 \times R/ \lambda$. With the pre-calibrated trusted loss inside the receiver, $\alpha$, detection efficiency of PDRMs can be calculated as $\eta'=\eta\times \alpha$.
4) Data acquisition and processing. In the experiment, AWG and DSO are working at trigger mode, and the frequency of the trigger signal is the same as the driving signal of AOMs. The driving signal of AOMs is used for controlling the on-off of the optical link, with the help of a power meter for monitoring the intensity of the optical signal, so quantum/reference signal and shot noise data can be separated in real time. Then, quantum/reference signal and shot noise data are captured by DSO, and they are processed by the same DSP algorithms for parameters estimation and raw  {data} achievement.

The DSP algorithms mainly include: 
 {1) Frequency offset estimation.} For Alice's and Bob's lasers, a frequency offset of about $1.55$~GHz is set to achieve the intermediate frequency signal and suppress low-frequency noises. However, due to the wavelength shift of the lasers, the accurate center frequency of quantum and reference signals is unknown. Therefore, frequency offset estimation in the frequency domain is performed. 
 {2) Bandpass filtering.} A frequency-domain  {bandpass} filter is used for quantum and reference signals, and the bandwidth is $1.3$~GHz and $200$~kHz, respectively. 
 {3) Frequency shift compensation.} A digital frequency shift with $750$~MHz is performed for the reference signal to remove the center frequency difference between the quantum and reference signal. 
 {4) Digital demodulation and carrier recovery.}
The X and P quadrature of quantum states are demodulated from the intermediate frequency signal digitally, and the carrier frequency shift and phase noise introduced by Alice's and Bob's lasers are compensated with the help of the high-power reference signal for the digital demodulation and carrier recovery. 
 {5) Matched filtering. }
An RRC filter with a roll-off factor of $0.3$ is used to filter the baseband quantum signal. 
 {6) Pilot-aided equalization.}
With the help of the time-domain superimposed training sequence and least-mean-square algorithm, a real-valued finite-impulse response filter is implemented. Details of the equalization are given in the supplementary materials. Here, the main functions of the real-valued finite-impulse response filter are X and P quadrature imbalance compensation, residual inter-symbol interference removal, and residual phase noise compensation. Notably, the procedure of equalization is realized by a fractional interval equalizer with $4$~times oversampling. After the above processing, the parameters estimation of the achieved raw key is performed to evaluate the performance of the network.

~\\
\textbf{Security analysis with one-time calibration}
~\\
 {In the scene where a coherent state is divided and sent to different Bobs, all Bob's responses contribute to the security of the entire network.}
The main task of security analysis is to estimate the secret key rate of the QKD links between Alice and different Bobs in the network. The end users are reasonably assumed to be trusted, where Eve has no access to the devices of the receiver or collaborates with Bobs.
The security analysis of each QKD link is based on the entanglement-based (EB) scheme, and the equivalence of EB and the prepare-and-measure (PM) scheme in practical experiment builds the basis of parameter estimation, including the replacement of the source, the shot noise unit (SNU) calibration, the practical detector module and the estimation of the covariance matrix.

The replacement of the source is the core of using an EB scheme to analyze a practical PM system. The source of the PM scheme in a practical system, which is the Gaussian modulated coherent state, is replaced by a two-mode squeezed vacuum (TMSV) state. With heterodyne detection on one mode of the TMSV state, the other mode is projected to a coherent state $(X_{B_0},P_{B_0})$ based on the heterodyne detection result $(X_{A_x}, P_{A_p})$.
Here, $(X_{B_0},P_{B_0})=k(X_{A_x}, P_{A_p})$, $k=\sqrt{2\times(V-1)/(V+1)}$, and $V=V_{mod}+1$. $V_{mod}$ is the modulation variance, and $(X_{B_0},P_{B_0})$ is the modulation data of the PM scheme, which can be calibrated in experiment. Therefore, with $(X_{B_0},P_{B_0})$ and $V_{mod}$, the source replacement is performed, to achieve the data of mode $A_x$ and $A_p$ in EB scheme. Since all Bobs in the network need to uniform the detection data, a one-time SNU calibration strategy is adopted to simplify the SNU calibration process by redefining the shot noise unit as the sum of the variances of shot noise and electronic noise. Using one-time SNU calibration, the electronic noise and the detection efficiency of a practical detector are both modeled as the loss. Since the electronic noise is not calibrated, the loss caused by the electronic noise is untrusted and the loss introduced by the limited detection efficiency is trusted. Based on the modulation data and the detection data of all end users in the PM scheme in experiment, we can estimate a covariance matrix which contains all the information for security analysis.


~\\
\textbf{Author contributions}
~\\
Y. Pan, Y. Bian, B. Xu and Y. Zhang conceived the idea. The experiments were conducted by Y. Pan with input from Y. Bian, H. Wang, Y. Pi, and J. Yang. Y. Bian conducted the  {protocol design} and simulations with input from Y. Pan, X. Xu, and J. Dou. L. Ma,  {Y. Luo} and Y. Li conducted the post-processing. Y. Pan and Y. Bian analyzed the data with input from Y. Li, W. Huang, S. Yu, Y. Zhang, and B. Xu. Y. Pan, Y. Bian, {S. Pirandola} and Y. Zhang prepared the manuscript with input from all authors. The work was contributed by Y. Pan and Y. Bian equally, and supervised by B. Xu and Y. Zhang.

~\\
\textbf{Acknowledgements}
~\\
This work was supported in part by the National Key Research and Development Program of China (Grant No. 2020YFA0309704), the National Natural Science Foundation of China (Grant Nos U22A2089, 62001044, 62101516, 62171418, 62201530 and 61901425), the Sichuan Science and Technology Program (Grant Nos 2023JDRC0017, 2023YFG0143, 2022ZDZX0009 and 2021YJ0313), the Natural Science Foundation of Sichuan Province (Grant Nos 2023NSFSC1387 and 2023NSFSC0449), the Basic Research Program of China (Grant No. JCKY2021210B059), the Equipment Advance Research Field Foundation (Grant No. 315067206), the Chengdu Key Research and Development Support Program (Grant No 2021-YF09-00116-GX).

\bibliography{apssamp}

\end{document}